\documentclass[manuscript]{aastex}

\shorttitle{GrayStar}
\shortauthors{Short}

\begin{document}


\title{GrayStar: A Web application for pedagogical stellar atmosphere and spectral line modelling 
and visualisation}


\author{C. Ian Short}
\affil{Department of Astronomy \& Physics and Institute for Computational Astrophysics, Saint Mary's University,
    Halifax, NS, Canada, B3H 3C3}
\email{ishort@ap.smu.ca}




\begin{abstract}

GrayStar is a stellar atmospheric and spectral line modelling, post-processing, and visualisation code, suitable for 
classroom demonstrations and laboratory-style assignments, that has been developed in Java and deployed in 
JavaScript and HTML.  The only software needed to compute models and post-processed
observables, and to visualise the resulting
atmospheric structure and observables, is a common Web browser.  Therefore, the code will run on any common PC or related X86 (-64) computer 
of the type that typically serves classroom data projectors, is found in undergraduate computer laboratories, 
or that students themselves own, including those with highly portable form-factors such as net-books and tablets.  
The user requires no experience
with compiling source code, reading data files, or using plotting packages.  More advanced students can
view the JavaScript source code using the developer tools provided by common Web browsers. 
 The code is based 
on the approximate gray atmospheric solution and runs quickly enough on current common PCs to provide 
near-instantaneous results,
allowing for real time exploration of parameter space.  
I describe the user interface and its inputs and outputs and suggest
specific pedagogical applications and projects.  Therefore, this paper may serve as a GrayStar user manual
for both instructors and students.
In an accompanying paper, I describe the
computational strategy and methodology as necessitated by Java and JavaScript.
I have made the application itself, and the HTML, CSS, JavaScript,
and Java source files available to the community.  
The Web application and source files may be found at {\url www.ap.smu.ca/$\sim$ishort/GrayStar}.

\end{abstract}



\section{Introduction}

\paragraph{}

 The stellar atmospheres and spectroscopy component of the essential undergraduate major or
honours astrophysics curriculum includes:
1) Relations among basic stellar parameters and overall radiation-related observables, as determined
for spherical emitters of black-body radiation, 2) The astrophysical basis for the relations between MK 
spectral class and effective temperature ($T_{\rm Eff}$), and MK
luminosity class and surface gravity ($\log g $), particularly the role of temperature, as 
determined by Saha-Boltzmann statistics for
excitation and ionization equilibrium, 3) The basic physics that determines average vertical atmospheric structure,
such as hydrostatic equilibrium (HSE) and the pressure equation of state (EOS), 4) The relation between atmospheric vertical temperature structure and
the distribution of the emergent surface intensity field in wavelength (spectral features) and in
angle of emergence with respect to the local surface normal (limb darkening), as determined approximately
by the LTE Eddington-Barbier relation, and 5) The relation between spectral line strength and shape, and the number
density of absorbers in the relevant atomic energy level ({\it ie.} the simple curve of growth (COG) of a spectral line).
These topics are crucially important for the astrophysical interpretation of spectra, not just from single stars, but also
from optically thick structures
generally, including interstellar medium (ISM) structures, disks on any scale, planetary atmospheres, and 
collections of spatially unresolved stars (integrated light populations), 
and are important for students going into many areas of astronomy and astrophysics.  

\paragraph{}

Because so many different physical laws and processes are involved in determining atmospheric structure and
the value of observables, the natural approach to pedagogy (and research!) is the parameter perturbation
analysis, in which we ask students to consider what might change if one stellar or spectral line formation
parameter is varied while the others are held fixed. 
The basic imaginary apparatus for these pedagogical thought experiments 
is the ``star with parameter knobs''.  
The purpose of GrayStar is to make this imaginary apparatus
a virtual apparatus, equipped with a virtual photometer, spectrophotometer, spectrograph, and optical 
interferometer, and to thus make the thought experiments simulated experiments. 
This would serve both for in-class demonstrations
that can form the basis of interactive pedagogy, and for laboratory-style assignments where students
investigate the dependence of structure and observables on parameters and explore relationships. 
 This would be pedagogically useful in courses ranging from first year undergraduate 
astronomy courses aimed at science majors, where one could demonstrate the parameter-dependence of 
the simplest observables, to introductory graduate courses, where one could explore the parameter dependence 
of the vertical atmospheric structure, demonstrate the implications of the Eddington-Barbier relation in detail,
or have the students study, adapt, and enhance the source code itself.           
 This is now more feasible than ever because commodity personal 
computers are now able to execute floating point operation- and
 memory-bandwidth- intensive operations quickly enough to execute 
 scientific computations that are not disk IO intensive, and are now suitable for execution of scientific 
codes at at least the pedagogical level of sophistication.

\paragraph{}

The pedagogical need for a virtual star with ``parameter knobs'' suggests that we look to atmospheric modelling and spectrum
synthesis codes and their suites of post-processing and visualisation tools. 
However, research-grade atmospheric modelling and spectrum synthesis codes, and their typical standard
 output files encoding observable data, are not necessarily suitable for wide pedagogical use for many
 reasons.  Because research-grade codes must read in large data files containing atomic and molecular 
line data and converge on a structure solution iteratively, the results are not available instantaneously
in a way that allows for real-time exploration with {\it ad hoc} parameters, even when 
run with their simplest level of realism.  The university-supplied computers that typically serve classroom data 
projectors and that are found in undergraduate computing labs, and the computers that students typically
 own themselves, normally have mass-market graphical-user-interface- (GUI-) based operating systems 
(OSes) that are not equipped with the tools and libraries needed to compile, link, and run complex 
codes written in typical scientific programming languages such as FORTRAN and C.  Neither are such
 computers particularly suitable for developing and running programs to read data files containing
 the results of pre-computed models and for plotting the contents thereof.  Many first and second year
 astronomy courses are taught by instructors ({\it eg.} part-time faculty, lecturers) who may not necessarily have
 the facility, nor the inclination, to obtain data files holding the results of research-grade atmospheric
 modelling and spectrum synthesis calculations, and to develop procedures for reading and plotting 
their contents, let alone to compile and run the codes themselves.  

\paragraph{}

 GrayStar is a simple stellar atmospheric
 and spectral line modelling, post-processing, and visualisation code that has been designed to be suitable for  
pedagogical use by instructors and students with no experience with producing an executable 
file from source code, or with producing routines to read data files or with data visualisation, and who 
have access only to computers with mass-market GUI-based OSes.  The atmospheric structure is computed
 using the approximate the gray solution (among other less crude approximations described below), obviating the need
 for input-output- (IO- ) intensive atomic and molecular ``big data'' handling and for iterative 
convergence.  The code is written in JavaScript, is processed by a Web browser's JavaScript interpreter and
the client's CPU, and 
displays its results in the browser window.  Therefore, it is certain to run successfully 
on any computer platform for which a common Web browser is available, 
 which includes all mass-market X86 (and X86-64) platforms and OSes of the type that serve classroom
 data projectors, are found in university computer labs, and that are owned by students and instructors,
 including those machines with highly portable form-factors such as tablets and net-books.  

\paragraph{}

In addition to making the GrayStar executable universally and freely available through the World Wide
 Web ({\url www.ap.smu.ca/$\sim$ishort/GrayStar}), I also disseminate the JavaScript and Hypertext Mark-up Language (HTML) source code to any who are interested in 
having their own local installation, or in developing the code further.  
I stress the broader significance that common PCs running common OSes are now powerful enough, and common Web-browsers are now sophisticated
enough interpreters, that the the realm of pedagogically useful scientific programming is now accessible in a framework that is free,
common, and allows both the application and its source code to be immediately shared over the Web.

\paragraph{}

In Section \ref{sUI} I describe the GrayStar user interface; 
in Section \ref{sCode} I briefly describe some of the special considerations relevant to developing and
deploying scientific modelling codes in Java and JavaScript; 
and in Section \ref{sApps} I describe a number of
 pedagogical demonstrations and lab exercises for which GrayStar is ideally suited.

\begin{figure}
\plotone{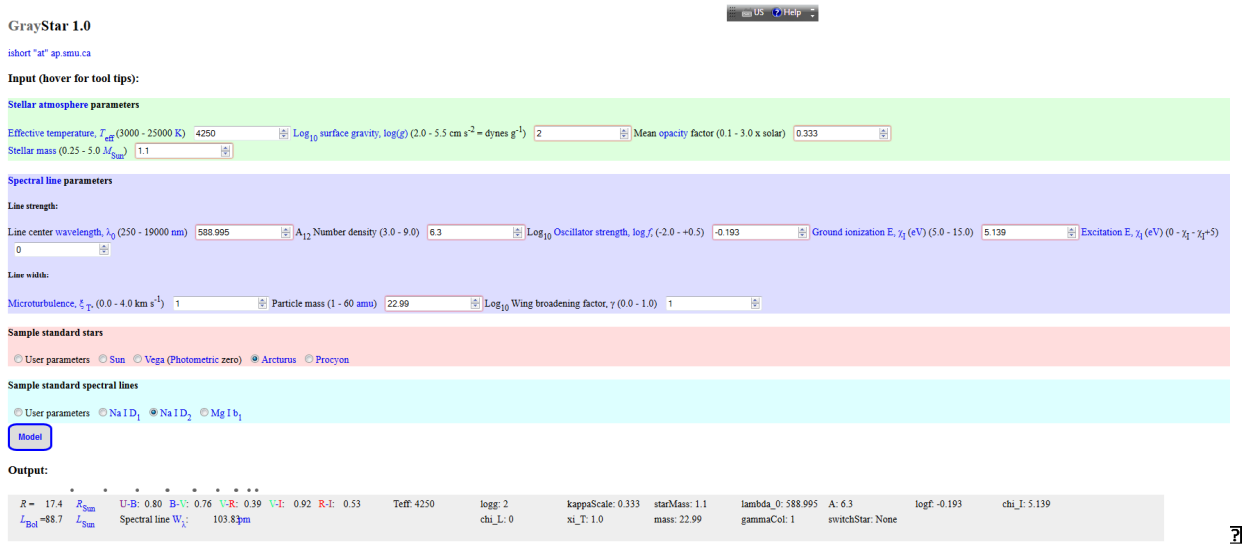}
\caption{A screen-shot of the input and textual output areas of the GrayStar GUI.
  \label{fGUIin}
}
\end{figure}

\begin{figure}
\plotone{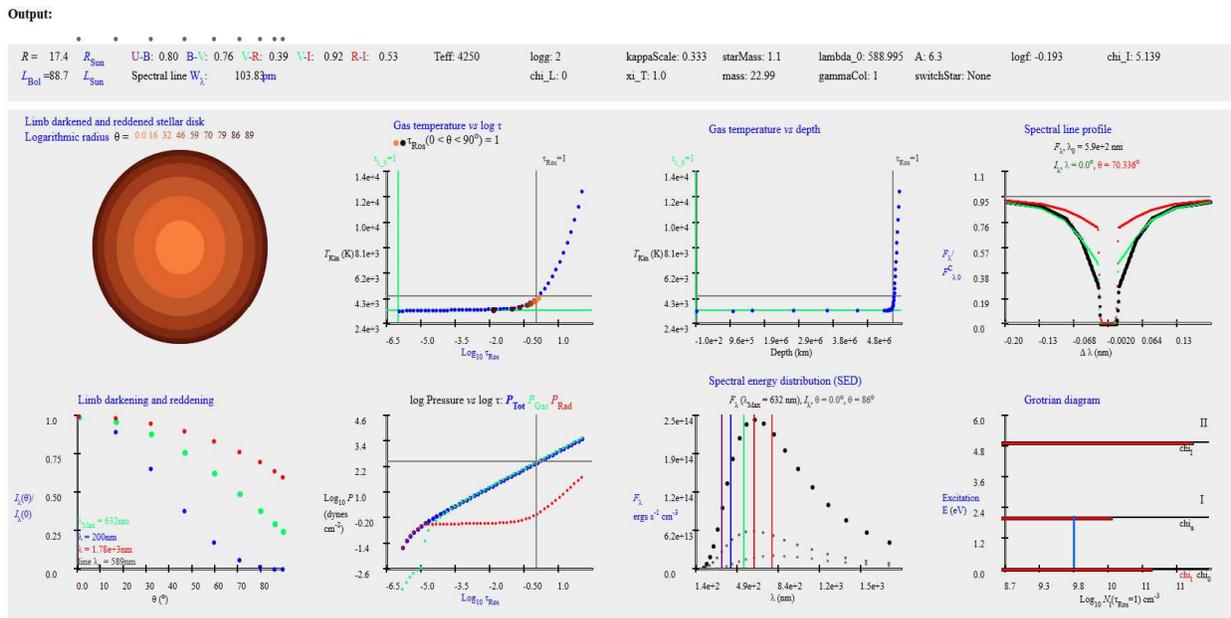}
\caption{A screen-shot of the output area of the GrayStar GUI showing the textual output section
and the eight plots of the graphical output section.
  \label{fGUIout}
}
\end{figure}


\section{User interface \label{sUI}}

\subsection{Input}

GrayStar presents the user with a browser window with 12 labeled 'text-box'-style
 input fields (Fig \ref{fGUIin}), four for stellar parameters and eight for spectral line parameters.
(Note that in
 the case of a gray atmospheric model, the microturbulence parameter, $\xi_{\rm T}$, is {\it only} a
 spectral line parameter, {\it not} an atmospheric parameter, because line extinction plays no role 
in determining the atmospheric structure.)  

\subsubsection{Stellar parameters }

1) Effective temperature, $T_{\rm Eff}$, in K, 2) Logarithmic surface gravity, $\log_{\rm 10} g$, 
in $\log$ cm s$^{-2}$ ($\log$ dynes g$^{-1}$), and 3) Multiplicative factor, $x$, for the Rosseland mean mass extinction
 co-efficient, $\kappa_{\rm Ros}$.  These are required to compute a model.  In addition, the interface expects 
4) Stellar mass, $M$, in solar units, $M_{\odot}$, for calculating the radius, $R$, and thus 
the bolometric luminosity, $L_{\rm Bol}$, in solar units, for purely pedagogical reasons.

\subsubsection{Spectral line parameters }

These are divided into two groups: those that mainly affect line {\it strength}, and those that mainly
affect line {\it width} or shape:

\paragraph{Line strength }
1) Line center wavelength, $\lambda_{\rm 0}$, in nm, 2) Logarithmic {\it total}
number density of the extinguishing species in the ``A$_{12}$'' system, $A = \log_{\rm 10} N/N_{\rm H} + 12$, 
3) Unit-less quantum
mechanical oscillator strength, $f_{\rm lu}$, of the corresponding bound-bound ({\it b-b}) atomic
transition, 4) Ground state ionization energy, $\chi_{\rm I}$, of the neutral ionization stage in eV, 5)
 Excitation energy, $\chi_{\rm l}$ of the lower atomic energy level of the corresponding {\it b-b}
transition, in eV, with respect to the ground state of the {\it neutral} stage.  If the value of
$\chi_{\rm l}$ (line parameter 5) is {\it less} than that of $\chi_{\rm I}$ (parameter 4), the spectral
 line corresponds to a {\it b-b} transition of the {\it neutral} ionization stage (I).  Otherwise, it
corresponds to a {\it b-b} transition of the {\it singly ionized} stage (II).

\paragraph{Line width and shape }

6) Mass of the absorbing species in atomic mass units (amu) (affects {\it thermal} core broadening), 
7) Microturbulent RMS velocity, $\xi_{\rm T}$, in km s$^{-1}$ (affects {\it non-thermal} core broadening),  
8) Logarithmic van der Waals damping enhancement factor, $\log_{\rm 10}\gamma$ (s$^{-1}$) 
(affects {\it wing} damping ).

\paragraph{}

Because this is a pedagogical application, guidance is provided for determining physically realistic
 and illustrative values.  The four stellar parameter input fields are pre-filled with default 
values for the Sun ($T_{\rm Eff}/\log g/x/M = 5778/4.44/1.0/1.0$), and the eight spectral line parameter 
fields are pre-filled with default values that yield a moderate spectral line ({\it ie.} with an 
approximately Gaussian profile) for a star with the Sun's stellar parameters.  The cgs system of units is used 
consistently, except for those values where standard practice is to use other units to ensure well-
normalized quantities (km s$^{-1}$ for $\xi_{\rm T}$, eV for atomic $\chi$ values).  
Moreover, the input fields are annotated with suggested 
ranges for the input values, and these ranges are enforced in the code itself to prevent students from
 inadvertently crashing the code by entering values that would lead to numerical pathologies.  I have
 also added ``tool tips'' to those field labels that are less self-explanatory that present additional information about the input 
parameters when the user hovers over a label, and most parameters labels are also linked to explanatory Web pages. 
If a user has their own installation of GrayStar, they can edit the links so that the fields point
to local on-line resources, thus embedding GrayStar in a local pedagogical framework.

\subsubsection{Pre-set models}
 
Finally, the user is given the option
of selecting from among four pre-set standard stars: the Sun (G2 V), Vega (A0 V, photometric zero), Arcturus (K1.5 III), 
and Procyon (F5 V-IV), and three pre-set Fraunhofer lines: The \ion{Na}{1} D$_{\rm 1}$ and D$_{\rm 2}$ lines, and the
 \ion{Mg}{1} b$_{\rm 1}$ line.  Comparison of the \ion{Na}{1} D$_{\rm 1}$ and D$_{\rm 2}$ lines demonstrates
COG effects in a multiplet (a doublet in this case), and the  \ion{Mg}{1} b$_{\rm 1}$ line demonstrates the
$T_{\rm Eff}$ variation of a line arising from an excited level.

\subsection{Output}

When the user runs a calculation by clicking the ``Model'' button, textual information and 
eight graphs are immediately displayed with the results of the calculation (Fig. \ref{fGUIout}):

\paragraph{Textual output: }

The values of the 12 input parameters are echoed back to the user.  This is important because 
the parameters used in the model may differ from those supplied if one or more parameters was
outside the ``safe'' range.  Any parameter that has been over-ridden is highlighted in red
to draw the student's attention.
The computed values of $R$ and $L_{\rm Bol}$ are displayed in solar units,  
along with five photometric colour indices in the Johnson-
Cousins  $UBV(RI)_{\rm C}$ system ($U-B$, $B-V$, $V-R$, $V-I$, $R-I$).  The colour indices are normalized 
with a single-point 
calibration to a GrayStar model computed with Vega's input parameters ($T_{\rm Eff}/log g/x = 9550/3.95/0.333$ \citep{castelli}).  
The equivalent width, $W_\lambda$, 
of the spectral line is displayed in pm (picometres).  The non-standard unit, pm (equal to 10 m\AA), was chosen for 
pedagogically-motivated consistency with the units of wavelength (nm).  

\paragraph{Graphical output: }

The GUI displays graphs of 
1) A physically based limb-darkened and limb-{\it reddened} rendering of the
projected stellar disk scaled logarithmically with radius, 
2) $T_{\rm Kin}$ {\it vs} logarithmic Rosseland optical depth, 
$\log\tau_{\rm Ros}$, 
3) Kinetic temperature, $T_{\rm Kin}$, in K {\it vs} 
geometric depth in km, 
4) logarithmic total pressure ($\log P_{\rm Tot}$), gas pressure 
($\log P_{\rm Gas}$), and bolometric radiation pressure ($\log P_{\rm Rad}$) {\it vs} 
$\log\tau_{\rm Ros}$, 
5) Limb darkening profiles, {$I_\lambda(\theta)\over I_\lambda(0)$ {\it vs} 
$\theta$ at the wavelength of maximum flux ($\lambda_{\rm Max}$), at representative continuum 
wavelengths in the near UV and near IR, and at spectral line center ($\lambda_{\rm 0}$), 
6) Surface flux 
spectral energy distribution (SED), $F_\lambda(\lambda)$ {\it vs} $\lambda$, and surface intensity 
SEDs, $I_\lambda(\lambda)$ at $\theta$ values of $\approx 0^\circ$ and $\approx 87^\circ$ for the 200
 (UV) to 2000 (IR) nm $\lambda$ range, 
7) the $F_\lambda(\lambda)$ spectral line profile and the
 $I_\lambda(\lambda)$ line profiles at  $\theta$ values of $\approx 60^\circ$ and $\approx 87^\circ$,
and 
8) A Grotrian diagram showing the atomic energy, $\chi_{\rm l}$, and logarithmic level population, $N_{\rm l}$,
at $\tau_{\rm Ros} = 1$ of four key $E$-levels: the lower and upper levels of the $b-b$ transition and the ground states of the
neutral and ionized stages, along with the transition itself.  

\paragraph{}

Plot 1) is labeled with the $\theta$ values of the emergent $I_\lambda(\theta)$ beams with respect
to the local surface-normal, color coded for consistency with the corresponding annuli in the 
limb-darkened, limb-reddened image. 
In plot 2) of $T_{\rm Kin}$ {\it vs} $\log\tau_{\rm Ros}$ the depths of $\tau_{\theta} = 1$ for
each $I_\lambda(\theta)$ beam sampling the radiation field in angle is shown with colour-coded symbols that
correspond to the limb- darkened and reddened rendering in plot 1).  Therefore, plots 1), 2), and 5) 
work together to provide a powerful, direct explanation of limb-darkening in terms of the LTE Eddington-Barbier relation.  
In the plots 2) through 4) displaying the atmospheric structure the depths where the continuum and line-center 
monochromatic optical depths scales are approximately unity (depths of $\tau_{\rm Ros}$ and $\tau_{\lambda 0} \approx 1.0$) 
are indicated.  
In plot 6) displaying the SEDs the central wavelengths of the 
$UBV(RI)_{\rm C}$ bands 
are indicated with appropriately colour-coded markers, and the value of ($\lambda_{\rm Max}$) is displayed. 

\section{Modelling \label{sCode}}

\paragraph{Java and JavaFX development }

 To take advantage of the robust development support provided by Integrated Development Environment 
applications (IDEs), I developed the code in Java, using version 1.8 of the Java Development Kit (JDK 1.8).  
Java has strong data typing and interface declaration rules,
 allowing the IDE to immediately catch bugs caused by most common coding errors as the code is being 
developed, including those arising from a mis-match between the types and numbers of the arguments and 
of the parameters of a function ({\it ie.} a Java ``method'').  Additionally the IDE provides the usual 
standard out (stdout) and standard error (stderr) channels for the Java interpreter/compiler to communicate 
with the developer, and provides robust error messaging.  All these are crucial to developing scientific 
computational codes of even moderate complexity.  This {\it development} version of the code uses the 
JavaFX library to provide the GUI, and has thus been named GrayFox, and has also been made available to 
the community through the GrayStar WWW site.  

\paragraph{JavaScript and HTML deployment }

Because the Java Run-time Environment (JRE) allows 
pre-compiled executable code to be downloaded and run on the client, and because Java has full 
file I/O capability, it poses a significant security threat.  Therefore, a difficult and financially 
expensive security protocol requires the deployer of Java codes to digitally authenticate their code 
with a certificate purchased from a trusted certificate issuer, and this poses a significant barrier 
to the free and wide dissemination of such codes in the academic sector.
To circumvent this difficulty of Java deployment, the code was ported to JavaScript and HTML for 
Web deployment.  With JavaScript applications, the client browser down-loads source code that can be visually 
inspected with typical browser developer tools, and JavaScript does not have file I/O capability.  
Therefore, the onerous and expensive 
authentication protocol is not required, and the code can be executed transparently by a client with 
typical default security settings.  
However, because JavaScript does not have IDE support, nor file I/O capability, 
it is more difficult to trouble-shoot and debug.  
Therefore, the recommended 
work-flow for further development is to develop the code in Java, then port it to JavaScript and HTML 
for deployment.  Porting the modelling algorithms is straightforward because, with the exception of variable and function 
(Java method) declarations, the syntax is identical (an exception is object declaration, but I 
have been unable to think of a way in which object oriented programming would benefit atmospheric 
modelling and spectrum synthesis!).  This {\it deployment} version is called GrayStar.      
Because Java and JavaScript development of modelling codes is not a strong part of the scientific programming culture,
and because I will make the source code publicly available for those who wish to understand and develop it, or
have their students study and modify it,
I provide a significant level of detail in an accompanying paper.  

\paragraph{HTML visualisation }

The biggest dis-incentive to scientific programming and visualisation with JavaScript and HTML is the need to
manually emulate the functionality of a plotting package ({\it egs.} IDL, gnuplot) starting from the primitive ability
of HTML to place a rectangle of given dimensions at a given location in the browser window, as specified in absolute device coordinates.
However, the code in the graphical output section of GrayStar may be taken as a template for how this problem can be
addressed, and adapted to other uses.

\paragraph{ }

GrayStar solves the static, $1D$ plane-parallel, horizontally homogeneous, local thermodynamic equilibrium 
(LTE), gray atmosphere problem, evaluates the formal solution of the radiative transfer equation to compute the
outgoing surface monochromatic specific intensity, $I_\lambda(\tau_\lambda=0, \theta)$,  and computes various
observables including the SED, photometric colour indices, and the profile of a representative spectral absorption line 
using the ``core plus wing'' approximation to a Voigt function profile.  (I note that all these approximations, 
{\it except the gray solution}, are not especially restrictive in the context of research-grade modelling!)  
The theoretical basis is taken 
entirely from \citet{rutten}.  The least realistic of these assumptions, by far, and the most expediting, 
is the gray solution, in which the monochromatic mass extinction co-efficient, $\kappa_\lambda(\lambda)$ 
is assumed to be constant as a function of wavelength, $\lambda$ (the Gray approximation), although it varies with depth,
and the angle-moments of the radiation field are related through the first and second Eddington approximations. 
 This yields an enormous simplification of the atmospheric structure problem in that the vertical kinetic 
temperature structure, $T_{\rm Kin}(\tau_{\rm Ros})$, can be calculated analytically, and the remaining structure variables 
({\it eg.} pressure ($P(\tau_{\rm Ros})$), density ($\rho(\tau_{\rm Ros})$)) can be calculated in a single pass with no need for iterative 
convergence.  The gray $T_{\rm Kin}(\tau)$ structure is most conceptually self-consistent at depth in the atmosphere when 
it is computed on the Rosseland optical depth scale, $\tau_{\rm Ros}$ ({\it ie.},  
$T_{\rm Kin}(\tau_{\rm Ros}))$.  Therefore, I set the gray value of $\kappa_\lambda$ at each depth to be equal (approximately) 
to the corresponding Rosseland mean mass extinction coefficient ({\it ie.} 
$\kappa_\lambda(\tau_{\rm Ros}, \lambda) = \kappa_{\rm Ros}(\tau_{\rm Ros})$) using the procedure
described in the accompanying paper on methods.
The numbers of points sampling the atmosphere in 
$\tau_{\rm Ros}$ (50), the radiation field in $\log\lambda$ (20) and $\theta$ (9), have been set to values close to 
the minimum that are useful to optimize the speed of execution.

  \section{Pedagogical applications \label{sApps}}

 It is worth noting that the GrayStar GUI is an HTML Web page like any other, and thus the usual methods for managing the display of content, and for capturing
textual content, are available: A presenter can enhance clarity by zooming, isolate areas of interest by re-sizing the browser, and show direct comparison
of output from runs with different parameters by launching multiple instances of the browser, or by using multiple browser tabs, each running 
GrayStar, and a student can capture textual output such as colour indices, $W_{\lambda}$ values, {\it etc.}, by cutting and pasting to a 
common spreadsheet program for analysis and plotting.  In particular, because GrayStar's graphical output consists of 
HTML instructions to the browser's rendering engine rather than pixelated bitmap information (such as that encoded in jpegs, gifs, {\it etc.}),
the graphics are scale-invariant and remain sharp at high zoom factors, which is an important consideration when presenting in large
classrooms.
The ability to conduct real-time numerical experiments with a simulated stellar atmosphere and spectrum in the classroom
is suitable for interactive pedagogical methods in which, for example, the instructor has the students predict the
outcome of a change in one or more parameters after discussing the situation among themselves.

\paragraph{}

The following pedagogical applications of GrayStar modelling can serve as demonstrations during lecture presentations, or as lab exercises. 
These are only the most obvious applications that suggest themselves immediately - other applications may occur with experience.  Plot numbers
refer to the enumeration given in Section \ref{sUI}.

\paragraph{}

1) Exploration of the variation of the value of peak spectral brightness, $\lambda_{\rm Max}$, and of the photometric colour indices with the value of $T_{\rm Eff}$, and 
comparison of the $T_{\rm Eff}$ variation of the different colour indices with each other (Plot 6), \newline

2) The role of 
$P_{\rm Rad}$ in supporting a stellar atmosphere, and its variation with stellar parameters (Plot 4), \newline

3) The role of $\kappa_{\rm Ros}$ in determining
the scale height (geometric thickness) of a stellar atmosphere (Plot 3),  \newline

4) Comparison of $F_\lambda$ and $I_\lambda(\theta)$ SEDs and line profiles and the relation to limb darkening, (Plots 6, 7, 5)\newline

5) Exploration of the monochromatic limb darkening,
$I_{\lambda}(\theta)$, with $\lambda$ and relation to the atmospheric vertical structure, $T_{\rm Kin}(\tau_{\rm Ros})$, and the variation of the
$T$ sensitivity of the Planck function, $B_\lambda(T, \lambda)$, with $\lambda$ (Plots 5, 2), \newline

6) Investigation of how the line core width varies with both the mass of the absorbing particle and the value of $\xi_{\rm T}$, and of how
 line damping wing strength varies with $\log g$ to illustrate pressure (collisional) broadening, \newline

7) Investigation of the curve of growth (COG, $W_\lambda(N_{\rm l}, f_{\rm lu})$) of a spectral line by varying both $A$ and $f_{\rm lu}$
throughout the range from a weak Gaussian line through  to a strong line to a saturated line with Lorentzian wings, (Plot 7) \newline

8) Comparison of the line strength ($W_\lambda$) variation with $T_{\rm Eff}$ 
among lines that belong to the neutral (I) and singly ionized (II) stages, and that arise from the ground ($\chi_l = 0$) or an excited
($\chi_l > 0$) atomic $E$-level .   \newline 

9) Investigation of the LTE Eddington-Barbier description of stellar spectral absorption line formation by comparison of the depths of 
$\tau_{\rm Ros}\approx 1$ and $\tau_{\lambda 0}\approx 1$ for lines of various strength, and the relation with the $T_{\rm Kin}(\tau_{\rm Ros})$ structure
(Plot 2), \newline

\paragraph{}

At the advanced undergraduate or graduate level, students can be asked to modify and add to the source code itself.  To modify and run the Java development version,
instructors and students need to download and install JDK 1.8 or later, and version 8.0 or later of the NetBeans IDE,
both available free of charge from Oracle's WWW site.
To trouble-shoot the JavaScript deployment version with diagnostic print statements (console.log()), the ``Developer tools'' accessible 
from the browser menu must be enabled, and the ``console'' selected. 

%
%
%

\section{Conclusions}

 GrayStar allows real-time exploration and investigation of stellar atmospheric structure and spectral
line profiles with ``on-demand'' parameters suitable for classroom demonstration and student laboratory 
assignments.  The JavaScript and HTML code is robustly platform-independent across all common types of
university-supplied and student-owned computers.  GrayStar allows pedagogical demonstration of 
most, if not all, major topics in the undergraduate astrophysics curriculum that are related to 
stellar atmospheres and spectra.  A local installation of GrayStar can be embedded in the local
pedagogical framework by editing the pedagogical links. 
 In addition to making GrayStar available for local installation and use by the astronomy teaching community,
I also encourage active development and adaptation of the code.  

\paragraph{}

  JavaScript, a language that can be interpreted by any common Web browser and executed on any common commodity 
personal computer for which a modern browser is available, is sophisticated enough as a programming language 
to allow development of scientific simulations at at least a pedagogically useful level of realism.  The ability of
JavaScript to manipulate HTML documents allows the results of simulations to be visualised in the Web browser.   
Commodity computers are now powerful enough to execute such JavaScript simulations instantaneously.   This allows 
``toy'' models of natural and physical systems to be simulated and visualised in a way that allows for 
pedagogical experimentation, classroom demonstration, and exploration of parameter space with no requirement
of computational or visualisation skills on the part of the instructor or student, and with no need
to install special purpose software.  Codes can be developed in Java, thus taking advantage of the powerful and mature 
developer support framework for Java, including IDEs, then straightforwardly ported to JavaScript.  

\paragraph{}

The biggest dis-incentive to scientific programming and visualisation with JavaScript and HTML is the need to 
manually emulate the functionality of plotting packages ({\it eg.} IDL) starting from the primitive ability
to place a rectangle of given dimensions at a given location in the browser window in absolute device coordinates.
However, the code in the graphical output section of GrayStar may be taken as a template for how this problem can be 
addressed, and adapted to other uses.

\paragraph{}

This opens 
up an entire vista of computational pedagogical possibilities, and the pedagogical stellar atmospheric and spectral line
modelling described here is only one example.  For example, within the field of stellar astrophysics, a similar approach could be taken for 
pedagogical simulation of stellar interior structure in the polytrope approximation, allowing classroom demonstration and
student laboratory investigation of the dependence of stellar structure and related observables on various
independent parameters - a ``stellar interior with knobs''. 
I stress again the broader significance that common PCs running common OSes are now powerful enough, and common Web-browsers are now sophisticated
enough interpreters, that the the realm of pedagogically useful scientific programming is now accessible in a framework that is free,
common, and allows both the application and its source code to be immediately shared over the Web.

\paragraph{}



\acknowledgments

\clearpage



\clearpage







\end{document}